\documentclass[11pt,a4paper]{article}

\usepackage[margin=1in]{geometry}
\usepackage{amsmath,amssymb}
\usepackage{booktabs}
\usepackage{hyperref}
\usepackage{authblk}
\usepackage[dvipsnames]{xcolor}
\usepackage{tcolorbox}
\usepackage{tabularx}
\usepackage{titlesec}
\usepackage{fancyhdr}
\usepackage{enumitem}
\usepackage{graphicx}
\usepackage{appendix}

\tcbuselibrary{breakable,skins}

\definecolor{boxbg}{HTML}{EFF6FF}
\definecolor{exbg}{HTML}{EFF6FF}

\newtcolorbox{calloutbox}[1]{
  enhanced,
  colback=boxbg, colframe=NavyBlue,
  colbacktitle=NavyBlue!15, coltitle=black,
  fonttitle=\bfseries, title={#1},
  boxrule=0.6pt, titlerule=0.3pt
}

\newtcolorbox{examplebox}[1]{
  enhanced,
  colback=exbg, colframe=NavyBlue,
  colbacktitle=NavyBlue!15, coltitle=black,
  fonttitle=\bfseries, title={#1},
  boxrule=0.6pt, titlerule=0.3pt
}

\newcolumntype{L}[1]{>{\raggedright\arraybackslash}p{#1}}
\newcolumntype{C}[1]{>{\centering\arraybackslash}p{#1}}

\titleformat{\section}{\large\bfseries}{\thesection.\quad}{0em}{}
\titleformat{\subsection}{\normalsize\bfseries}{\thesubsection\quad}{0em}{}

\hypersetup{colorlinks=true,linkcolor=NavyBlue,urlcolor=NavyBlue,citecolor=NavyBlue}

\newcommand{\appendixsectionformat}{%
  \titleformat{\section}{\large\bfseries}{Appendix~\thesection.\quad}{0em}{}%
}

\title{\LARGE The Weak Signal Cultivation Model\\[6pt]
\large\textit{A Human-Centric Framework for Frontline Risk Detection,\\
Signal Tracking, and Proactive Organizational Resilience}\\[8pt]
\normalsize\textsc{White Paper}}

\author[1]{Maurice Codourey}
\author[2]{Emmanuel A.\ Gonzalez}
\affil[1]{\small codourey@proton.me}
\affil[2]{\small emmgon@gmail.com}
\date{2026}

\begin{document}
\maketitle

\begin{abstract}
\noindent This white paper introduces the Weak Signal Cultivation Model (WSCM). WSCM is a human-centric framework for detecting, structuring, and tracking weak risk signals as observed by frontline staff. The model centers on a continuous $[0, 10] \times [0, 10]$ coordinate field---the Weak Signal Cultivation Field, in which each identified signal is positioned as a node on two independent dimensions: its current Risk Intensity ($x$) and its Risk Growth Potential ($y$). Represented as a \textbf{risk locus}, nodes move across the field over time as new team assessments or measurements arrive. The locus reflects the signal's trajectory across four possible regions: Question Marks, Lit Fuses, Sleeping Cats, and Owls. Through this graphical approach, bridging risk communication from the frontline experience to management decision-making is made through a single organizational vocabulary. The model introduced in this document is designed to serve as a practitioner tool and a conceptual foundation for AI-supported analytics.
\end{abstract}


\vspace{10pt}
\begin{center}
\small
\begin{tabularx}{\textwidth}{L{1.8cm}X}
\toprule
\multicolumn{2}{l}{\textbf{Abbreviations and Notation}} \\
\midrule
\multicolumn{2}{l}{\textit{Acronyms}} \\
\addlinespace
BCG & Boston Consulting Group (Growth-Share Matrix) \\
CSCW & Computer-Supported Cooperative Work \\
HCI & Human-Computer Interaction \\
NRS & Numeric Rating Scale (0--4, used for coordinate elicitation) \\
SMS & Safety Management System \\
SSI & Session Severity Index \\
WSCM & Weak Signal Cultivation Model \\
WSFQ & Weak Signal Farming Quadrant (predecessor framework) \\
\addlinespace
\multicolumn{2}{l}{\textit{Key symbols (full notation in Table~5)}} \\
\addlinespace
$x, y$ & Risk Intensity and Risk Growth Potential coordinates $[0, 10]$ \\
$\tau$ & Dimensionless time: elapsed days divided by reference period $T_{\text{ref}}$ \\
$\alpha, \beta$ & Recency weights for $x$ and $y$ updates \\
$\kappa$ & Consensus momentum (directional agreement amplifier) \\
$\mu$ & Passive $y$-decay rate (default 0.087; half-life $\approx$ 8 reference periods) \\
$d$ & Euclidean distance from origin; SMS escalation at $d \geq 7.07$ \\
$S$ & Session Severity Index (reporting metric; does not drive position) \\
$f$ & Frequency of occurrence count within current observation year \\
\bottomrule
\end{tabularx}
\end{center}
\vspace{4pt}

\section{Introduction}

Organizations in safety-critical industries expend significant effort and resources in deploying formal risk management systems. And yet the simple pattern behind major incidents has been repeatedly observed, over and over, in accident reports and investigations: before the accident occurred, warning signs were present and observed by people close to the work, but never reached the threshold of a record or high enough visibility that might set in motion a response. The signals were weak (ambiguous, isolated, or easy to explain away), and the systems intended to capture risk were not designed to capture them.

These shortcomings have been extensively studied, and named, in the safety science literature. In rich, case-study based catalogs, Kletz and Amyotte (2019) assembled decades of industrial disasters that had occurred in the face of identifiable precursors that were never acknowledged. Reason (1997) described organizational accidents as the confluence of latent conditions that each appeared benign on their own. Dekker (2014) contended that human error is typically a symptom of system-wide vulnerabilities rather than an individual failure. The unifying insight in these works is simple: the information to prevent the event was available, but the infrastructure to surface it, develop it, and act on it was not.

Current efforts to address this gap tend to one of two divergent poles. Technology-heavy solutions including digital near-miss reporting systems, AI-powered anomaly detectors, and enterprise risk platforms offer significant analytic power, but their infrastructure requirements, training needs, and engagement overhead present barriers to practical implementation on many frontlines (Pfeiffer et al., 2010; Aboagye-Nimo et al., 2015). Informal approaches including safety walks, open debriefs, and conversations capture information rich in experiential knowledge (Eraut, 2004) but leave no structured record, and have no capability for tracking changes in signal characteristics over time.

The Weak Signal Cultivation Model (WSCM) is a response to this divide. It is a low-technology, structured tool that any team can adopt with a whiteboard, a printed form, or a spreadsheet. It provides weak signals a visual home, a position on a shared field, and a mechanism for recording changes in that position over time. It requires no specialized software to use, but the outputs are directly usable in AI-supported analytics when that integration is eventually sought.

This document has three closely related contributions:

\begin{enumerate}
\item The Weak Signal Cultivation Field is a new continuous $[0, 10] \times [0, 10]$ coordinate framework, conceptually inspired by the BCG Growth-Share Matrix (Henderson, 1970), for plotting and visualizing weak signals relative to two independent risk dimensions.
\item The concept of \textbf{risk locus}, that is, the path that a weak signal's node traces in the field across time. This is used mainly as an analytical tool where the path bears the history of information about the signal over time.
\item The WSCM's position at the boundary between current practice and future capability: Teams can use it today with nothing more than a printed copy and a writing instrument, and the coordinate structure it produces is ready for direct ingestion by AI-supported analytics tools when they become available.
\end{enumerate}

The choice of the term \textit{cultivation} in the model name is intentional. Detection is a one-time event; monitoring is passive observation; tracking is following something already in motion. Cultivation is different because it's an active ongoing engagement with something whose nature isn't yet clear, across multiple work sessions, until it either develops into a distinct concern or fades away into irrelevance. The agricultural metaphor comes from the original name of the framework on which this work is built (Weak Signal Farming Quadrant), and it expresses what the model actually requires of teams: plant the observations in a shared field and revisit them often to see what grows.

\begin{figure}[b]
\begin{calloutbox}{Remark. On the origin of this work}
The Weak Signal Cultivation Model evolved from the Weak Signal Farming Quadrant (WSFQ), first conceived and applied by Maurice Codourey in 2019 in healthcare settings, where the need to surface subtle, context-dependent risk signals from frontline nursing and clinical staff motivated the original framework design (Codourey, 2025). The model has since been extended and refined through interdisciplinary exchanges and exploratory sessions in industrial safety contexts. This paper constitutes the first formal publication of the WSCM.
\end{calloutbox}
\end{figure}

\section{Background and Related Work}

\subsection{Weak Signals and the Challenge of Early Detection}

In the management literature, the challenges of interpreting early warnings about environmental change are discussed under the concept of weak signals (WS). Originating in Ansoff (1975), the term points to the frequently observed phenomenon that, in cases where the environment has changed suddenly and profoundly, the organization had at hand incomplete and ambiguous early signs of what was to come, but had no means of responding.

In safety science, the same observation applies: weak signals are latent risk or near-failure phenomena that require structured attention to interpret and develop. Near-misses, behavioral aberrations, and other subtle changes in the operational processes are in principle information carriers about future failure. Near-misses are perhaps the most well-established in this respect in the safety literature (Gnoni \& Saleh, 2017; Van der Schaaf, 1992). These are incidents that could have resulted in harm but did not, and are therefore assumed to have the same causal makeup as adverse events (Hollnagel et al., 2006).

At the same time, as a typical WS, they are weak in the eyes of traditional risk approaches, due to their ambiguity, rarity, or any other quality which makes them difficult to quantify. The case study in Farinha, Vesnic-Alujevic and P\'{o}lvora (2023) on current signal detection practices, shows that current approaches often break down for this very reason: questionnaire-based collection leaves the signal unstructured and unenriched, too-early handing over to IT systems breaks down context before collective sense-making has a chance, and lack of continued group engagement degrades the quality of the interpretations over time. The analysis suggests that participatory and iterative practices are more effective, and this is the design principle of the WSCM.

\subsection{Safety-II and the Value of Everyday Knowledge}

Modern safety science has moved from Safety-I (the study and prevention of failures) to Safety-II, which is the study and facilitation of the conditions under which everyday work goes right (Hollnagel, 2014). In the Safety-II paradigm, frontline workers are seen as active knowledge sources whose adaptive workarounds are a wellspring of organizational intelligence not reflected in formal knowledge repositories. The hazard detection and assessment that frontline workers are constantly doing in their immediate environment has been examined through modeling work on hazard identification effectiveness in safety-critical industrial settings. That work suggests that even experienced and expert workers can systematically fail to notice hazards that are present but not salient (Gonzalez et al., 2015). The WSCM is designed to address exactly this gap.

The WSCM puts this viewpoint into practice by elevating frontline observation to primary data source status. Near misses, behavioral red flags, workflow disruptions, and low-level environmental changes are examples of the operational anomalies that seldom get reported. They are the WSCM's raw material. The quadrant field is not a top-down risk register that is imposed on frontline workers; it is a structured canvas into which workers' observations are cast, from which patterns are extracted, and which are collectively interpreted.

\subsection{Behavioral Nudging and Collective Habit Formation}

Thaler and Sunstein (2021) demonstrated that small and carefully designed interventions into a physical or social environment can predictably shape collective behavior without resorting to mandates or penalties. The WSCM structural features incorporate this insight. The regular session rhythm, the shared visual field, and the named signal lexicon all amount to a gentle nudge in the direction of constant risk awareness. Teams who use the quadrant regularly are not simply gathering data; they are forming a shared attentional habit. Weak signal detection becomes a practiced organizational capability instead of a periodic exercise.

UNICEF's handbook \textit{Human-Centred Design in the Field} (2019) has captured this dynamic in human development contexts, using case studies to show that small, persistent interventions such as marginal changes to service delivery, community relations, or product design can shape decision-making more robustly than large-scale messaging campaigns. The biweekly session rhythm of the WSCM was designed to have this effect.

\subsection{Participatory Design as Iterative Sense-Making}

Hasdell (2016) has shown that participatory design processes based on iterating cycles of ideation, observation, and group reflection are more useful than a priori frameworks for recording messy, emergent phenomena. Collective sense-making, how groups construct shared understanding from ambiguous cues (Weick, 1995), is a core element of this dynamic: weak signals become organizational signals through structured group interpretation, not individual judgment. The Cynefin framework (Snowden \& Boone, 2007), for example, with its categories of simple, complicated, complex and chaotic organizational contexts, suggests that weak signals belong in the complex domain: where cause and effect can only be discerned in hindsight and where probe-sense-respond methods outperform analyze-then-act methods. The WSCM is explicitly designed as a probe-and-sense instrument for the complex domain.

HCI and CSCW research has explored how collaborative sense-making practices can be supported through structured tools during ambiguous, high-stakes situations. Varanasi et al.\ (2023) propose design principles for crisis sense-making platforms. They find that resilience results from patterned collective interpretation, not centralized interpretation, a concept the WSCM directly instantiates at the team level. Poon et al.\ (2023) explore computer-mediated sharing circles used by home care workers. They find that structured peer-to-peer communication reveals operational concerns not captured by formal channels, an effect the WSCM's cultivation session is designed to achieve. These results are consistent with broader CSCW findings that how we design collective tools shapes what kinds of knowledge an organization can generate (Bjørn \& Østerlund, 2014).

Wicht (2025) has pushed this argument further, theorizing that weak signals are more than early warnings; they are prompts to imagine different organizational futures. Reading them requires the challenge-dominant-narrative orientation to speculation that the WSCM's open session structure is meant to cultivate.

\subsection{Existing Risk Tools and the Gap the WSCM Fills}

Several established tools serve in the same function space as the WSCM, but none of them solves the problem in question. Standard risk matrices, the near-ubiquitous likelihood-by-severity grids from virtually every SMS, sort and categorize named, mature risks into fixed buckets. They are good for risks we already know, but provide no structure to capture the vague, pre-quantitative signals front-line workers face every day. An unnamed risk cannot be on a risk matrix. Vaughan (1996) describes this mode of failure in her analysis of the Challenger disaster: the signals were there in the data, but successively discounted in a process Vaughan calls the ``normalization of deviance'', whereby anomalous signals are encountered so often they come to be accepted as normal. The WSCM's cultivation field is designed to avert this kind of drift: by keeping every signal in view and in an observable position, normalization becomes more difficult to achieve.

Near-miss reporting systems (Van der Schaaf, 1992) aim squarely at the capture problem, soliciting incident reports from frontline workers. These systems are event-driven rather than signal-driven: they track what happened rather than what might be happening. A near-miss report is a single frame; the WSCM tracks trajectories. These reports are typically aggregated into databases and parsed by safety professionals months later; they do not support real-time collective sense-making as the WSCM's session structure does.

The BCG Growth-Share Matrix (Henderson, 1970), the structural inspiration for the WSCM's quadrant orientation, shows that 2D coordinate fields with named regions can become powerful shared vocabularies for organizational decision-making. In this way, the BCG matrix arguably succeeded in giving managers a shared language for portfolio strategy. The WSCM takes this insight and applies it to risk: the four regions (Question Marks, Lit Fuses, Sleeping Cats, Owls) have a vocabulary role for safety teams similar to that of Question Marks, Cash Cows, Stars and Dogs for business strategists.

Bow-Tie models and fault-tree analyses enable rigorous causal reasoning about known failure modes, but require specialist expertise to construct and maintain. These are analytical tools for mature and known risks; they are not signal detection tools for the unknown and emerging. The WSCM is meant to complement, not replace, such approaches, operating upstream in the risk lifecycle by surfacing the weak signals that might one day become candidates for Bow-Tie analysis.

\section{The Weak Signal Cultivation Model}

\subsection{Base Model}

The WSCM maps weak signals onto a two-dimensional coordinate system, the Weak Signal Cultivation Field. Each signal is represented by a node at location $(x, y)$, where:

\begin{itemize}
\item $x$ --- \textbf{Risk Intensity}: the current size or expected impact of the signal. $x \in [0, 10]$. $x = 0$ means that the signal carries no discernible risk; $x = 10$ means the signal is assessed to be as intense as possible.
\item $y$ --- \textbf{Risk Growth Potential}: the expected trajectory or growth rate of the signal. $y \in [0, 10]$. $y = 0$ means that the signal is receding or thought not to be growing; $y = 10$ means the signal is assessed to be growing as quickly as possible.
\end{itemize}

The origin $(0, 0)$ represents zero risk on both dimensions. The far corner $(10, 10)$ represents maximum concern on both. The space is continuous, bounded but not discrete. Each real-valued point in the area $[0, 10] \times [0, 10]$ is a valid location.

The two axes are mathematically uncoupled: there are no cross-terms in the update equations, and a change in $x$ has no direct effect on $y$ and vice versa. It is possible for a signal to have high intensity but low growth potential (a big but currently stable risk) or low intensity but high growth potential (a currently small but fast-growing signal). This decoupling is the key design decision that enables the model to simultaneously capture what a risk is and where it is headed. We do not, a priori, know whether the two dimensions are also empirically uncoupled, whether in real data from organizations the members of a team who are asked to assess a set of weak signals assess the intensity and growth potential as two genuinely independent factors. This is an empirical question that field experiments will need to test.

The field is further subdivided into four named areas by the boundaries that meet at the field midpoint $(5, 5)$. These regions are described in Table~\ref{tab:regions}. The boundaries are pedagogical, not mechanical: no transition is triggered by crossing a boundary. The team only notes the signal's current quadrant membership.

\begin{table}[t]
\centering
\caption{The Four Regions of the Weak Signal Cultivation Field.}
\label{tab:regions}
\small
\begin{tabularx}{\textwidth}{|X|X|}
\hline
\multicolumn{2}{|c|}{\textit{High Risk Growth Potential ($y \geq 5$)}} \\
\hline
\textbf{Sleeping Cats} & \textbf{Owls} \\
\textit{(Dormant / Legacy Risks)} & \textit{(Managed, Active Risks)} \\
$x < 5 \cdot y \geq 5 \cdot$ Low intensity, high growth & $x \geq 5 \cdot y \geq 5 \cdot$ High intensity, high growth \\[4pt]
\textit{Weak but accelerating---dormant risk awakening. Requires sustained vigilance.} &
\textit{Escalating and actively managed---full organizational attention required.} \\
\hline
$\leftarrow$ \textit{Lower Risk Intensity ($x < 5$)} & \textit{Higher Risk Intensity ($x \geq 5$)} $\rightarrow$ \\
\hline
\textbf{? Question Marks} & \textbf{Lit Fuses} \\
\textit{(Emerging / Weak Signals)} & \textit{(Hot Incident Candidates)} \\
$x < 5 \cdot y < 5 \cdot$ Low intensity, low growth & $x \geq 5 \cdot y < 5 \cdot$ High intensity, low growth \\[4pt]
\textit{Emerging weak signal---early stage, unclear, requires observation.} &
\textit{High intensity but growth still low---significant, localized concern.} \\
\hline
\multicolumn{2}{|c|}{\textit{Low Risk Growth Potential ($y < 5$)}} \\
\hline
\end{tabularx}
\end{table}

\begin{figure}[t]
\centering
\includegraphics[width=0.85\textwidth]{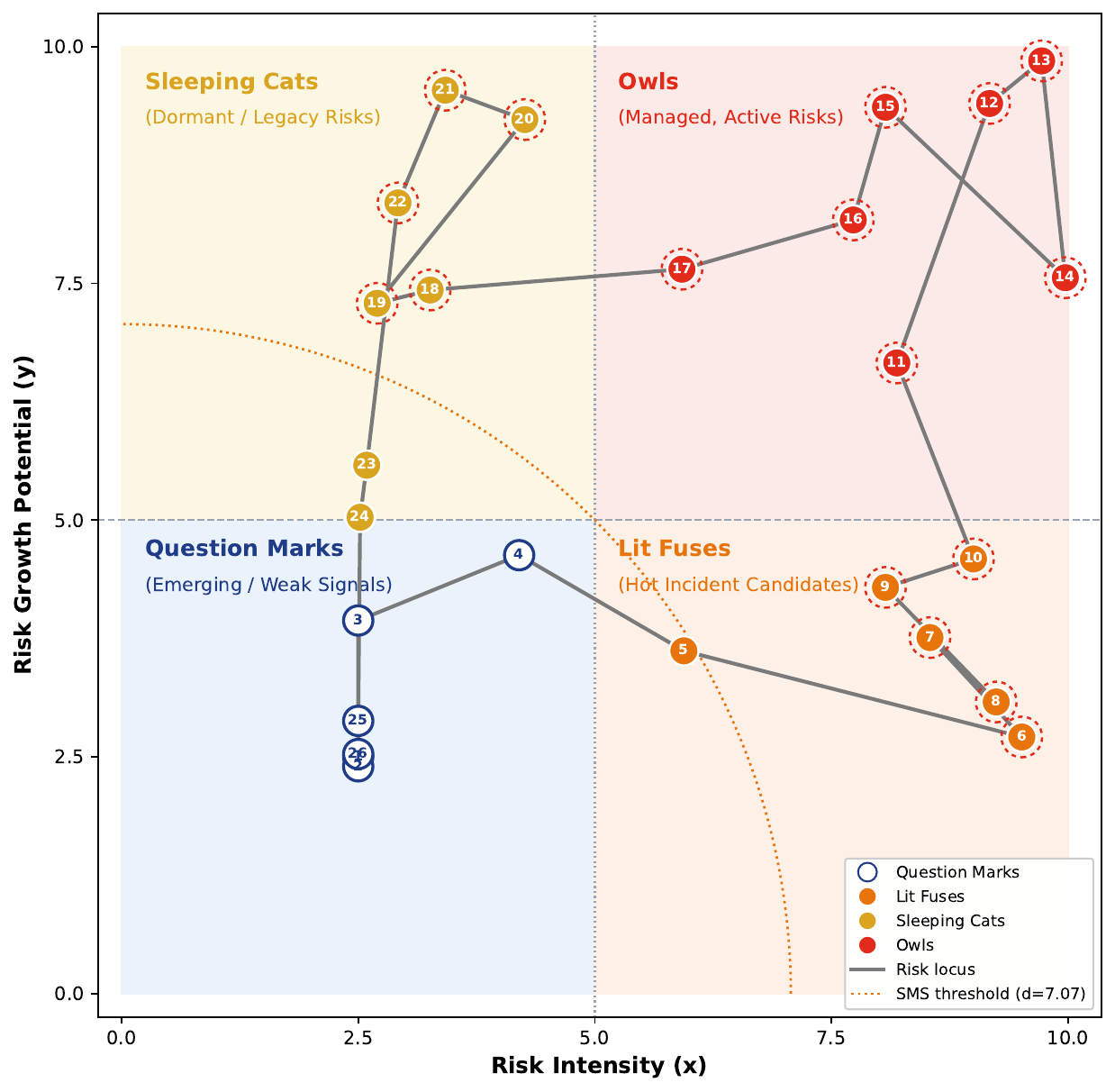}
\caption{Exemplary movement of the risk signal ``gas fumes'' through the Weak Signal Cultivation Field across 26 cultivation sessions (Table~\ref{tab:gasfumes}). The risk locus illustrates the typical risk progression: Question Marks (S1--S4) $\rightarrow$ Lit Fuses (S5--S10) $\rightarrow$ Owls (S11--S17) $\rightarrow$ Sleeping Cats (S18--S24) $\rightarrow$ Question Marks (S25--S26). Numbers represent sequential session observations. The dotted arc marks the SMS escalation threshold at $d = 7.07$; dashed rings mark sessions with SMS escalation active ($d \geq 7.07$).}
\label{fig:locus}
\end{figure}

A node does not jump from one category to another. It moves continuously across the field as new assessments come in. The current region only indicates its current location. A signal heading from Question Marks toward Lit Fuses is more worrying than one at rest in Lit Fuses but trending back toward Question Marks, even if their current positions in the field are similar. Direction of motion is as important as position. Figure~\ref{fig:locus} shows an example signal path across all four quadrants.

\subsection{The Risk Locus}

The most analytically important feature of the WSCM is the path a node has traced across the field over time, rather than its position at any single session. We call this path the \textbf{risk locus}, the continuous trajectory connecting successive node positions, indexed by the real calendar time at which each observation was made.

The risk locus, illustrated for the gas fumes signal in Figure~\ref{fig:locus}, makes visible what a static risk register cannot. It reveals the dynamics of a risk's evolution. A risk locus that moves quickly rightward and upward across the field signals acceleration toward crisis. A locus that bends back from Owls toward Lit Fuses after an intervention demonstrates measurable organizational response. A locus that oscillates within a small region of the field for many months reveals a chronic, managed tension that may warrant structural rather than tactical attention.

These are qualitatively different organizational situations that static risk scores systematically conflate. The risk locus preserves and communicates this difference in a form that is immediately interpretable by practitioners without statistical training.

\begin{figure}[b]
\begin{calloutbox}{Remark. The risk locus and AI analytics}
In future implementations, the risk locus---as a time-series of $(x, y)$ coordinate pairs---becomes the primary input to AI-supported pattern recognition systems. Machine learning models trained on locus shapes across multiple signals and organizations can identify characteristic escalation patterns, early warning trajectories, and intervention effectiveness signatures. The WSCM is designed from the outset so that what human teams produce manually is directly compatible with this analytical layer. The practitioner tool and the AI analytics share the same data structure.
\end{calloutbox}
\end{figure}

\subsection{The Four Regions in Detail}

Table~\ref{tab:regions-detail} expands each of the four regions with its coordinate boundaries and the practitioner response it calls for.

\begin{table}[t]
\centering
\caption{The four regions of the Weak Signal Cultivation Field with coordinate boundaries and descriptions.}
\label{tab:regions-detail}
\small
\begin{tabularx}{\textwidth}{L{2.5cm}L{10.5cm}}
\toprule
\textbf{Region} & \textbf{Description} \\
\midrule
Question Marks\newline$(x < 5, y < 5)$ &
Question Marks hold signals that are emerging, unclear, and at an early conceptual stage. Both intensity and growth potential are below the field midpoint. The signal has not yet established measurable momentum on either dimension. These are the `les petits riens' of organizational life: tiny observations that on their own say little, but together suggest something that is emerging worth watching. The right reaction is careful tending: name the signal, describe it sharply, start watching it regularly, and don't let the organization make you ignore it because you can't measure it yet. Many major incidents, in retrospect, spent time here.\\
\addlinespace
Lit Fuses\newline$(x \geq 5, y < 5)$ &
Lit Fuses hold signals with significant current intensity. They have crossed the $x = 5$ threshold, but their growth potential remains below the midpoint. The signal is present and impactful but has not yet fully escalated. This is the region of growing concern and proactive mitigation need. Teams encountering a signal in Lit Fuses should treat it with the seriousness of a formal incident report, not because harm has occurred, but because intensity is already high and trajectory could shift.\\
\addlinespace
Owls\newline$(x \geq 5, y \geq 5)$ &
Owls represent signals that are both intense and escalating. These are risks that have crossed both thresholds and now command full organizational attention. They are actively monitored and managed. The owl metaphor captures the character of this region: vigilant, patient, and nocturnal. A signal in Owls is not resolved; it is supervised. The underlying conditions that generated it have not been eliminated, only contained. The focus here is continuous management and containment.\\
\addlinespace
Sleeping Cats\newline$(x < 5, y \geq 5)$ &
Sleeping Cats hold residual or legacy risks. Current intensity is below the midpoint. The immediate danger appears to have receded, but growth potential is high, meaning the underlying conditions for reactivation are active. These signals are weak but accelerating or spreading. The sleeping cat metaphor is precise: the animal appears inert but retains the capacity to act without warning. Organizational memory about what previously activated these signals is a critical safeguard against their return.\\
\bottomrule
\end{tabularx}
\end{table}

\subsection{Positioning a Signal: The NRS Elicitation Protocol}

Placing a signal at coordinates $(x, y)$ requires no specialist measurement. It only requires structured team conversation, applied consistently at each cultivation session. Team members assess each active signal using a five-point Numeric Rating Scale (NRS, 0--4):

\begin{table}[t]
\centering
\caption{NRS elicitation questions for WSCM coordinate placement.}
\small
\begin{tabular}{L{1.5cm}L{8cm}C{2cm}}
\toprule
\textbf{Coord.} & \textbf{Elicitation question} & \textbf{Scale} \\
\midrule
$x$ & How severe or impactful is this anomaly right now? & NRS 0--4 \\
$y$ & How strongly is this anomaly growing or escalating? & NRS 0--4 \\
\bottomrule
\end{tabular}
\end{table}

\begin{table}[t]
\centering
\caption{NRS value to model coordinate conversion: $x_{\text{model}} = 2.5 \times \text{NRS}$.}
\small
\begin{tabular}{C{2cm}C{2.5cm}L{6cm}}
\toprule
\textbf{NRS value} & \textbf{Model value} & \textbf{Interpretation} \\
\midrule
0 & 0.0 & No concern---signal not present \\
1 & 2.5 & Weak signal---maximum entry level \\
2 & 5.0 & Boundary zone---moderate concern \\
3 & 7.5 & Significant concern \\
4 & 10.0 & Maximum concern \\
\bottomrule
\end{tabular}
\end{table}

\textbf{Entry constraint.} A signal can only enter the WSCM when every individual NRS score is $\leq 1$ on both axes. This constrains the initial position to $[0, 2.5] \times [0, 2.5]$---entirely within Question Marks. The entry constraint enforces the model's purpose: only genuine weak signals enter here. Mature, known risks with established organizational records do not enter via this pathway.

\textbf{Committee aggregation.} Each participating team member provides independent NRS scores for both $x$ and $y$. The session coordinate is the arithmetic mean of the $n$ assessors' scores, scaled to the model field:

\begin{equation}
x_{\text{new}} = 2.5 \cdot \frac{1}{n}\sum_{i=1}^{n} \text{NRS}_{x,i} \qquad\qquad y_{\text{new}} = 2.5 \cdot \frac{1}{n}\sum_{i=1}^{n} \text{NRS}_{y,i}
\end{equation}

The conversation that produces these scores matters as much as the numbers that result---it is where collective sense-making about organizational risk takes shape and gets updated (Weick, 1995). Consensus among many reporters is captured downstream in the momentum mechanism (Section~3.5).

\subsection{Coordinate Derivation: The Computational Model}

The WSCM uses a time-adaptive update system to move each signal's node as new session assessments arrive. Position is not recomputed from scratch each session. Instead, it is updated from the prior position through recency-weighted blending, so that recent observations carry more influence than older ones, but no single session dominates signal history. The steps below show summaries with default values as seen in Table~\ref{tab:params}. Appendix~A describes how each default value was determined.

\begin{table}[t]
\centering
\caption{Notation for WSCM coordinate update formulas.}
\small
\begin{tabularx}{\textwidth}{L{2cm}L{11cm}}
\toprule
\textbf{Symbol} & \textbf{Definition} \\
\midrule
$x, y$ & Current coordinates, Risk Intensity and Risk Growth Potential ($[0, 10]$) \\
$x_{\text{new}}, y_{\text{new}}$ & NRS-derived assessments for the current session \\
$\tau$ & Dimensionless time: $\tau = \Delta t / T_{\text{ref}}$ \\
$T_{\text{ref}}$ & Organizational reference period in days (default: 14) \\
$\alpha_{\text{base}}, \beta_{\text{base}}$ & Max recency weight ceilings for $x$ and $y$ (default: 0.90) \\
$\lambda, \nu$ & Time-sensitivity of $x$ and $y$ recency saturation (default: 0.75) \\
$\mu$ & Passive $y$-decay rate (default: 0.087; half-life $\approx$ 8 reference periods) \\
$y_{\min}$ & Floor preventing full $y$-decay while signal is open (default: 0.5) \\
$\delta, \eta, \phi$ & Consensus momentum weights: magnitude (0.5), sustained (0.3), volume (0.3) \\
$\psi$ & Reversal amplifier strength (default: 0.5) \\
$k_{\text{ref}}$ & Reference session count for full momentum effect (default: 5) \\
$n_{\text{ref}}$ & Reference reporter count for full volume effect (default: 5) \\
$\alpha_{\min}$ & Minimum $n$-cap floor for small committees (default: 0.7) \\
$d_{\text{close}}$ & Minimum distance from origin before closure is permitted (default: 1.0) \\
$f$ & Frequency of occurrence count within the current observation year \\
\bottomrule
\end{tabularx}
\end{table}

\textbf{Step 1 --- Dimensionless Time.} Raw elapsed time is converted to a dimensionless parameter $\tau$ that makes all update formulas portable across session cadences. $\Delta t$ is the number of calendar days elapsed since the previous session, and $T_{\text{ref}}$ is the organizational reference period (default: 14 days, matching a biweekly cadence):
\begin{equation}
\tau = \frac{\Delta t}{T_{\text{ref}}}
\end{equation}
At biweekly cadence, a session held on schedule always produces $\tau = 1.0$. A session held after a one-week gap produces $\tau = 0.5$; after a four-week gap, $\tau = 2.0$. Only $T_{\text{ref}}$ changes when deploying to a new organizational context.

\textbf{Step 2 --- Base Recency Weights.} New assessment information is weighted by elapsed time using exponential saturation functions. Sessions after a long gap carry higher update weight; sessions held very soon after the previous carry lower weight, guarding against overreaction to noise:
\begin{equation}
\alpha(\tau) = \alpha_{\text{base}} \cdot \left(1 - e^{-\lambda\tau}\right)
\end{equation}
\begin{equation}
\beta(\tau) = \beta_{\text{base}} \cdot \left(1 - e^{-\nu\tau}\right)
\end{equation}
At $\tau = 1$: weight $\approx$ 53\% of ceiling. At $\tau = 2$: $\approx$ 78\%. At $\tau = 3$: $\approx$ 90\%. The ceilings $\alpha_{\text{base}}$ and $\beta_{\text{base}}$ (both defaulting to 0.90) cap how much influence any single session can exert on $x$ and $y$ respectively, regardless of elapsed time. The $x$ and $y$ axes use independent time-sensitivity parameters ($\lambda$ and $\nu$) because intensity and growth potential respond differently to organizational dynamics.

\textbf{Step 3 --- Consensus Momentum ($\kappa$).} When a team consistently agrees and moves in the same direction across successive sessions, the model amplifies the update weight to reflect sustained directional consensus. $\kappa$ answers: should we move faster because multiple people have been consistently pushing in the same direction?
\begin{equation}
\kappa(\tau, k) = \rho \cdot \min\!\left(1,\;
\underbrace{\delta \cdot \frac{|x_{\text{new}} - x|}{10} \cdot \frac{k+1}{k_{\text{ref}}}}_{\text{magnitude--momentum}}
+ \underbrace{\eta \cdot \frac{k}{k_{\text{ref}}}}_{\text{persistence}}
+ \underbrace{\phi \cdot n_{\text{boost}}}_{\text{committee size}}\right)
\end{equation}
where $k$ is the count of consecutive sessions in the same direction (resets to 0 on reversal), $n_{\text{boost}} = \min(1, n/n_{\text{ref}})$ captures reporter volume, and the three terms inside $\min$ capture: magnitude amplified by momentum; pure sustained agreement; and reporter volume. The weighting coefficients are $\delta = 0.50$ (magnitude sensitivity), $\eta = 0.30$ (sustained agreement sensitivity), and $\phi = 0.30$ (reporter volume sensitivity). $k_{\text{ref}} = 5$ is the reference session count at which the momentum effect reaches full strength, and $n_{\text{ref}} = 5$ is the corresponding reference reporter count. A direction reversal is detected when $\operatorname{sign}(x_{\text{new}} - x) \neq \operatorname{sign}(x_{\text{new,prev}} - x_{\text{prev}})$; the same test applies independently to the $y$ axis.

The leading multiplier $\rho$ is a reversal amplifier controlled by $\psi = 0.50$:
\begin{equation}
\rho = \begin{cases} 1 + \psi \cdot n_{\text{boost}} & \text{if direction reversal and } n \geq n_{\text{ref}} \\ 1 & \text{otherwise} \end{cases}
\end{equation}
In the normal (non-reversal) case $\rho = 1$ and has no effect. When a large committee reverses its directional assessment, $\rho$ scales up $\kappa$ by up to 1.5$\times$, allowing the model to reposition the signal faster in response to a credible change of direction.

The $n$-dependent effective weight ceiling prevents small committees from driving large updates. $\alpha_{\min} = 0.70$ sets the floor: even a single assessor can move the node, but not as far as a full committee:
\begin{equation}
n_{\text{cap}} = \alpha_{\min} + (1 - \alpha_{\min}) \cdot \min\left(1, \frac{n}{n_{\text{ref}}}\right)
\end{equation}
\begin{equation}
\alpha_{\text{eff}} = \min(n_{\text{cap}},\; \alpha(\tau) + \kappa_x)
\end{equation}
\begin{equation}
\beta_{\text{eff}} = \min(n_{\text{cap}},\; \beta(\tau) + \kappa_y)
\end{equation}

\textbf{Step 4 --- Position Update.} When a session produces new assessments, $x$ is updated directly and $y$ incorporates a passive decay term reflecting the natural diminishment of escalation urgency when no new reports reinforce it:
\begin{equation}
x' = x + \alpha_{\text{eff}} \cdot (x_{\text{new}} - x)
\end{equation}
\begin{equation}
y_{\text{decay}} = y \cdot e^{-\mu\tau}
\end{equation}
\begin{equation}
y' = \max\!\left(y_{\min},\; y_{\text{decay}} + \beta_{\text{eff}} \cdot (y_{\text{new}} - y_{\text{decay}})\right)
\end{equation}

This asymmetry is by design. Intensity ($x$) does not decay without evidence. If a hazard was severe, it remains severe until someone reports otherwise. Growth potential ($y$) represents escalation trajectory, which naturally diminishes over time without fresh signals. The decay rate $\mu = 0.087$ gives a half-life of approximately 8 reference periods, meaning that $y$ halves roughly every 8 reference periods (about 16 weeks at biweekly cadence) if no new observations reinforce it. The floor $y_{\min} = 0.50$ prevents a signal from silently decaying to zero while it remains in the active register.

In the case where there is no new NRS assessment, Equations 10 to 12 will not apply, and the equations reduce into a pure passive decay expressed in Equation 13.
\begin{equation}
x' = x \qquad\qquad y' = \max\!\left(y_{\min},\; y \cdot e^{-\mu\tau}\right)
\end{equation}

\textbf{Step 5 --- Exit Threshold and SMS Escalation.} After every position update, the model checks the Euclidean distance from the origin:
\begin{equation}
d = \sqrt{x'^{2} + y'^{2}}
\end{equation}
When $d \geq 7.07$ ($= \sqrt{50}$, the distance from origin to field centre $(5, 5)$), the signal is escalated to the formal Safety Management System (SMS). The WSCM continues recording the risk locus after escalation for audit and organizational learning. Escalation is a parallel flag---not an exit from the model.

\textbf{Step 6 --- Session Severity Index (S).} After each coordinate update, a Session Severity Index is computed as a reporting and escalation-monitoring metric. $S$ does not drive coordinate position. The update equations above fully determine node movement. $S$ summarizes the combined weight of position, exposure, and occurrence frequency into a single number for prioritization dashboards and session-over-session trend tracking:
\begin{equation}
S = \frac{d}{d_{\max}} \cdot \ln(1 + f) \qquad\text{where}\quad d_{\max} = \sqrt{200} \approx 14.14
\end{equation}
Here $f$ is the cumulative frequency of occurrence count within the current observation year, recorded by the facilitator at each session. The logarithm provides diminishing returns: the jump from $f = 1$ to $f = 5$ contributes more to $S$ than the jump from $f = 15$ to $f = 19$, preventing high-frequency but low-concern signals from dominating the index.

\begin{table}[t]
\centering
\caption{Session Severity Index interpretation and recommended response.}
\small
\begin{tabular}{L{2.5cm}L{2.5cm}L{7cm}}
\toprule
\textbf{$S$ range} & \textbf{Status} & \textbf{Recommended action} \\
\midrule
$[0.0,\,0.5)$ & Low & Routine monitoring; log and observe \\
$[0.5,\,1.5)$ & Moderate & Team review recommended \\
$[1.5,\,2.5)$ & Elevated & Management notification; increase session frequency \\
$\geq 2.5$ & Critical & Escalation imminent or SMS already triggered \\
\bottomrule
\end{tabular}
\end{table}

\begin{table}[t]
\centering
\caption{Complete parameter reference with defaults for biweekly deployment. See Appendix~A for the determination basis and mathematical derivations.}
\label{tab:params}
\small
\begin{tabularx}{\textwidth}{L{2cm}C{2cm}L{9cm}}
\toprule
\textbf{Parameter} & \textbf{Default} & \textbf{Role} \\
\midrule
$T_{\text{ref}}$ & 14 days & Reference period; change only this for a new session cadence \\
$\alpha_{\text{base}}$ & 0.90 & Max recency weight ceiling for $x$; lower = more stability \\
$\beta_{\text{base}}$ & 0.90 & Max recency weight ceiling for $y$; lower = more stability \\
$\lambda$ & 0.75 & Time-sensitivity of $x$ recency saturation \\
$\nu$ & 0.75 & Time-sensitivity of $y$ recency saturation (independent of $\lambda$) \\
$\mu$ & 0.087 & $y$ decay rate; $y$ halves after $\approx$8 reference periods \\
$\delta$ & 0.50 & Magnitude sensitivity in $\kappa$ \\
$\eta$ & 0.30 & Sustained agreement sensitivity in $\kappa$ \\
$\phi$ & 0.30 & Reporter volume sensitivity in $\kappa$ \\
$\psi$ & 0.50 & Reversal amplifier strength \\
$k_{\text{ref}}$ & 5 & Reference session count for full momentum effect \\
$n_{\text{ref}}$ & 5 & Reference reporter count for full volume effect \\
$\alpha_{\min}$ & 0.70 & Minimum $n$-cap floor for small committees \\
$y_{\min}$ & 0.50 & $y$-floor; signal never silently disappears below this value \\
$d_{\text{close}}$ & 1.0 & Minimum distance from origin before closure is permitted \\
\bottomrule
\end{tabularx}
\end{table}

\begin{figure}[b]
\begin{calloutbox}{Remark. Formula Evolution}
The computational model presented here is the current best-practice specification for WSCM coordinate elicitation, position updating, and severity reporting. Parameter defaults were calibrated from simulation data across two worked signals: Gas Fumes (Section~5 of this document) and Complacency (Codourey, 2025). As noted in Section~7.2, a separate methods paper will extend and test these expressions across multi-site deployments. Readers should treat this formalization as an initial working specification. The core conceptual architecture---the continuous $[0, 10] \times [0, 10]$ coordinate field, the independence of $x$ and $y$ axes, and the risk locus as the primary analytical unit---remains stable regardless of future parameter refinements.
\end{calloutbox}
\end{figure}

\section{Applying the Model: The Session Structure}

\subsection{The Cultivation Session}

The WSCM is applied through regular cultivation sessions---structured team meetings, typically 30 to 60 minutes, in which active signals are reviewed, new observations are contributed, node positions are updated, and the risk locus of each signal is examined for directional change. The recommended cadence is biweekly, though organizations with high operational tempo may benefit from weekly sessions and those with lower activity may find monthly sessions sufficient.

A cultivation session has four components:

\begin{enumerate}
\item \textbf{Signal review} --- revisit each active signal. Has anything changed since the last session? New near-misses, behavioral observations, operational changes?
\item \textbf{Position update} --- for each signal with new observations, update the $(x, y)$ coordinates using the NRS inputs. Record the new position and add it to the risk locus.
\item \textbf{New signal intake} --- any observations from the period that do not yet have a signal entry? An isolated anomaly becomes a registered signal when the team believes it represents a pattern worth watching. If it meets the entry threshold, place an initial coordinate and begin the risk locus.
\item \textbf{Action decisions} --- for signals in Lit Fuses or moving rapidly in that direction, what is the team's response? Assign responsibilities, set a review date, and document the decision.
\end{enumerate}

\begin{figure}[b]
\begin{calloutbox}{Quick Start: Running Your First Cultivation Session}
Four prerequisites: (1) a team of four to eight people who work close to the operations you want to monitor; (2) a printed session form or whiteboard with the WSCM field drawn on it; (3) a facilitator willing to keep the meeting to thirty minutes; and (4) at least one observation that someone on the team believes is worth watching. At the first session, register one to three signals, assign each an NRS score on both axes, and place them on the field. That is your starting point. Return in two weeks and see what has changed.
\end{calloutbox}
\end{figure}

\subsection{Signal Registration}

Not every observation becomes a registered signal. The threshold for registration is low but not absent: an observation becomes a signal when at least one team member believes it represents a pattern worth watching, something seen more than once, or seen once in a context suggesting it could recur. This threshold is human and subjective by design. The organizational cost of registering an anomaly that turns out to be insignificant is negligible. The cost of failing to register one that matters is potentially catastrophic.

When a signal is registered, it is assigned three attributes by the team: a name, a definition and an initial position. The name should be something memorable, a word or short phrase that can be used by all members of the team. The definition needs to be clear enough that two independent observers would agree whether or not they had observed the signal. The initial position represents the best estimate of the team as to where the signal lies on the NRS inputs at the time of registration.

\subsection{Graduating and Retiring Signals}

Signals are not permanent. A signal that has spent many consecutive sessions in Sleeping Cats with no directional movement, confirmed effective controls, and consistent low-intensity readings may be retired from active tracking---but retained in the organizational memory record. Retiring a signal is a conscious team decision, not an automatic process. The record of its risk locus remains accessible.

Signals are never deleted. The organizational memory function of the WSCM depends on retaining the full history of every signal ever registered, including those that proved insignificant and those that materialized into incidents. Both categories contain learning.

\subsection{Implementation Considerations}

The session structure described above assumes a team that is already using the WSCM. This section addresses the practical questions that arise before and during early adoption: who should facilitate, what failure modes to expect, how long before useful signal history accumulates, and how to handle disagreement within the team.

\textbf{Facilitation.} The WSCM does not require a dedicated safety officer or specialist facilitator. A team lead or supervisor familiar with the operational context can run cultivation sessions effectively, provided they understand the four session components and the NRS observation inputs. The facilitator's primary role is process management, keeping the session within its time boundary, ensuring all active signals are reviewed, and preventing the session from drifting into incident investigation or root cause analysis, which are separate processes. The facilitation burden decreases as teams develop familiarity with the format; experienced teams typically complete signal review and position updates for ten to fifteen active signals within thirty minutes.

\textbf{Anticipated Failure Modes.} From the model's design, four potential failure modes should be considered in early deployment.

\textit{Anomaly inflation}---teams register every passing observation, leading to a field crowded with low-relevance entries that dilute attention from genuine weak signals. The registration threshold described in Section~4.2 is designed to prevent this. Facilitators should hold the threshold consistently in early sessions.

\textit{Convergence bias}---team members defer to the most senior voice when assessing input values, producing artificial consensus that suppresses genuine variation in observation (Tversky \& Kahneman, 1974). This can be mitigated by collecting individual assessments of the NRS inputs independently before the session and using the range of responses as discussion material rather than treating agreement as the default goal.

\textit{Drift toward incident management}---sessions begin to focus on explaining past events rather than detecting emerging signals. A simple session norm---``we are looking forward, not backward''---combined with a time-boxed agenda helps maintain the prospective orientation.

\textit{Abandonment after dormancy}---when no signals escalate over several consecutive sessions, teams may perceive the tool as generating no value and discontinue use. Consistent positioning in Sleeping Cats or Owls is itself organizationally meaningful data, confirming that controls are holding and no new emergence is occurring. Framing quiet sessions as confirmatory rather than empty supports sustained engagement.

\textbf{Time to Useful Signal History.} A single cultivation session produces a snapshot; it does not produce signal history. Organizationally actionable risk locus data (trajectories long enough to distinguish genuine escalation from noise) typically requires a minimum of six to eight sessions per signal, corresponding to approximately three to four months at biweekly cadence. Teams should treat the first eight weeks of WSCM deployment as a calibration period in which the primary output is team familiarity with the model's vocabulary and process.

\textbf{Handling Disagreement.} Disagreement among the team members over the correct $(x, y)$ position of a signal is information to be leveraged, not hidden. The process should be to record the spread of the assessments and the consensus position, identify where the disagreement is coming from, and use any lingering disagreement as a symptom that the definition of that signal needs to be clarified, or that the signal actually includes two separate things that should be tracked as separate signals. The WSCM is a conversation, not a vote. Productive disagreement is a part of that conversation, not a breakdown.

\textbf{Minimum Viable Deployment.} The WSCM can be used with one single operational team of four to eight people, a printed copy of the session form, and a facilitator willing to run a half hour meeting every two weeks. No digital supporting infrastructure is needed at this scale. Organizations deploying at the team level but wanting to span multiple teams should synchronize on a session form and signal naming conventions prior to roll out, as the comparability of the results across teams relies on the definitions being the same.

\section{Illustrative Application: The Risk Journey of Gas Fumes}

The following example (visualized in Figure~\ref{fig:locus}) traces the lifecycle of a single weak signal, intermittent gas fumes reported near a loading dock, across 26 biweekly cultivation sessions spanning 252 days (36 weeks). The data are author-constructed to demonstrate the model's behavior across all four field regions; they are not derived from field observation. Default model parameters apply throughout: $T_{\text{ref}} = 14$, $\alpha_{\text{base}} = \beta_{\text{base}} = 0.90$, $\lambda = \nu = 0.75$, $\mu = 0.087$. The example illustrates how the risk locus concept makes organizational response visible.

\textbf{Signal definition:} intermittent detection of gas odor near loading dock B, reported by multiple shift workers across different weeks. Scope: loading dock B and adjacent storage corridor.

\textbf{Session 1 --- Entry Calculation.} The signal enters on Day~0. Entry requires NRS $\leq 1$ on both axes. One assessor: NRS$_x = 1$, NRS$_y = 1$.
\[
x_1 = 2.5 \times 1 = 2.50 \qquad y_1 = 2.5 \times 1 = 2.50
\]
Initial placement: $(2.50, 2.50)$---positioned in Question Marks ($x < 5$, $y < 5$). The signal is weak on both dimensions, consistent with a first unconfirmed report.

\textbf{Session 2 --- Worked Update (Day 14, $\tau = 1.0$).} One assessor returns the same NRS values: $x_{\text{new}} = 2.50$, $y_{\text{new}} = 2.50$.

Recency weights (Equations 3--4):
\[
\alpha(1.0) = 0.90 \times (1 - e^{-0.75 \times 1.0}) = 0.475 \qquad \beta(1.0) = 0.475
\]

No momentum ($k = 0$); $n_{\text{boost}} = \min(1, 1/5) = 0.20$; $n_{\text{cap}} = 0.70 + 0.30 \times 0.20 = 0.76$. Momentum contribution: $\kappa = \phi \cdot n_{\text{boost}} = 0.30 \times 0.20 = 0.060$. $\alpha_{\text{eff}} = \min(0.76, 0.475 + 0.060) = 0.535$.

Position update (Equations 10--12):
\[
x_2 = 2.50 + 0.535 \times (2.50 - 2.50) = 2.500
\]
\[
y_{\text{decay}} = 2.50 \times e^{-0.087} = 2.292 \qquad y_2 = \max(0.5,\; 2.292 + 0.535 \times (2.50 - 2.292)) = 2.403
\]

Placement: $(2.500, 2.403)$ --- Question Marks.

\begin{table}[t]
\centering
\caption{Gas Fumes signal: 26-session risk locus across all four field regions. Asterisk ($*$) indicates SMS escalation active ($d \geq 7.07$).}
\label{tab:gasfumes}
\footnotesize
\begin{tabularx}{\textwidth}{C{0.6cm}C{0.8cm}C{1.0cm}C{1.0cm}L{1.8cm}C{0.5cm}L{5.5cm}}
\toprule
\textbf{S} & \textbf{Day} & $\mathbf{x'}$ & $\mathbf{y'}$ & \textbf{Region} & $\mathbf{f}$ & \textbf{Key event} \\
\midrule
1 & 0 & 2.50 & 2.50 & Q.\ Marks & 3 & Entry. Faint smell, reported 3 times before registration. \\
2 & 14 & 2.50 & 2.40 & Q.\ Marks & 4 & One new detection; no escalation. \\
3 & 28 & 2.50 & 3.94 & Q.\ Marks & 5 & Second reporter; $y$ rising. \\
4 & 42 & 4.20 & 4.63 & Q.\ Marks & 7 & Two more detections; approaching boundary. \\
5 & 49 & 5.94 & 3.62 & Lit Fuses & 9 & Frequent detections; crosses $x = 5$. \\
6 & 63 & 9.51 & 2.71 & Lit Fuses$^*$ & 12 & Peak intensity; SMS triggered ($d = 9.89$). \\
7 & 70 & 8.54 & 3.76 & Lit Fuses$^*$ & 15 & Intensity oscillating; $y$ beginning to rise. \\
8 & 77 & 9.24 & 3.08 & Lit Fuses$^*$ & 19 & High intensity; growth suppressed. \\
9 & 91 & 8.07 & 4.29 & Lit Fuses$^*$ & 23 & Sustained concern; $y$ creeping up. \\
10 & 98 & 9.00 & 4.59 & Lit Fuses$^*$ & 27 & $y$ approaching midpoint. \\
11 & 105 & 8.19 & 6.66 & Owls$^*$ & 33 & $y$ crosses 5; enters Owls. Management engaged. \\
12 & 112 & 9.17 & 9.40 & Owls$^*$ & 40 & Emergency response; $d = 13.13$. \\
13 & 119 & 9.72 & 9.85 & Owls$^*$ & 48 & Peak session: maximum risk locus point $(9.72, 9.85)$, $d = 13.84$. \\
14 & 133 & 9.97 & 7.56 & Owls$^*$ & 52 & $y$ declining; mitigation taking effect. \\
15 & 140 & 8.07 & 9.36 & Owls$^*$ & 54 & Intensity dropping; growth rebounds. \\
16 & 147 & 7.73 & 8.17 & Owls$^*$ & 55 & Both axes declining. \\
17 & 154 & 5.92 & 7.65 & Owls$^*$ & 56 & $x$ falling sharply; source identified. \\
18 & 161 & 3.26 & 7.43 & Sl.\ Cats$^*$ & 57 & $x$ crosses 5 downward; enters Sleeping Cats. \\
19 & 168 & 2.70 & 7.29 & Sl.\ Cats$^*$ & 57 & Intensity falling; growth still elevated. \\
20 & 182 & 4.26 & 9.23 & Sl.\ Cats$^*$ & 58 & $y$ near peak in Sleeping Cats. \\
21 & 189 & 3.42 & 9.54 & Sl.\ Cats$^*$ & 58 & Controls applied; intensity steady. \\
22 & 196 & 2.92 & 8.35 & Sl.\ Cats$^*$ & 58 & $y$ declining. \\
23 & 210 & 2.59 & 5.58 & Sl.\ Cats & 58 & Growth potential decaying. \\
24 & 224 & 2.52 & 5.03 & Sl.\ Cats & 58 & $d = 5.63$; approaches safe zone. \\
25 & 238 & 2.50 & 2.88 & Q.\ Marks & 58 & Returns to Question Marks. \\
26 & 252 & 2.50 & 2.53 & Q.\ Marks & 58 & De-escalated; retirement candidate. \\
\bottomrule
\end{tabularx}
\end{table}

Several features of this trajectory deserve attention. First, the signal spent four sessions in Question Marks (S1--S4) before a two-reporter assessment at Session~5 pushed intensity across $x = 5$ into Lit Fuses. This delay is correct---the model requires accumulating evidence before escalating, preventing premature organizational response to an unconfirmed report.

Second, the SMS escalation threshold ($d \geq 7.07$) was triggered at Session~6 and remained continuously active through Session~22, then dropped below threshold from Session~23 onward as the signal de-escalated. This extended escalation window covers the Lit Fuses, Owls, and early Sleeping Cats phases, showing that the distance-based threshold captures sustained organizational risk beyond peak intensity alone.

Third, the signal entered Sleeping Cats at Session~18 when $x$ fell below~5 following source identification and control implementation. However, $y$ remained elevated through the early Sleeping Cats sessions---the underlying growth potential had not yet resolved. The sleeping cat metaphor fits: the immediate hazard had receded, but the conditions for reactivation persisted. As clean readings accumulated, the growth potential finally decayed; the signal returned to Question Marks by Session~25 and closed Session~26 at $(2.50, 2.53)$, a de-escalated retirement candidate retained in the register.

\begin{figure}[t]
\centering
\includegraphics[width=0.95\textwidth]{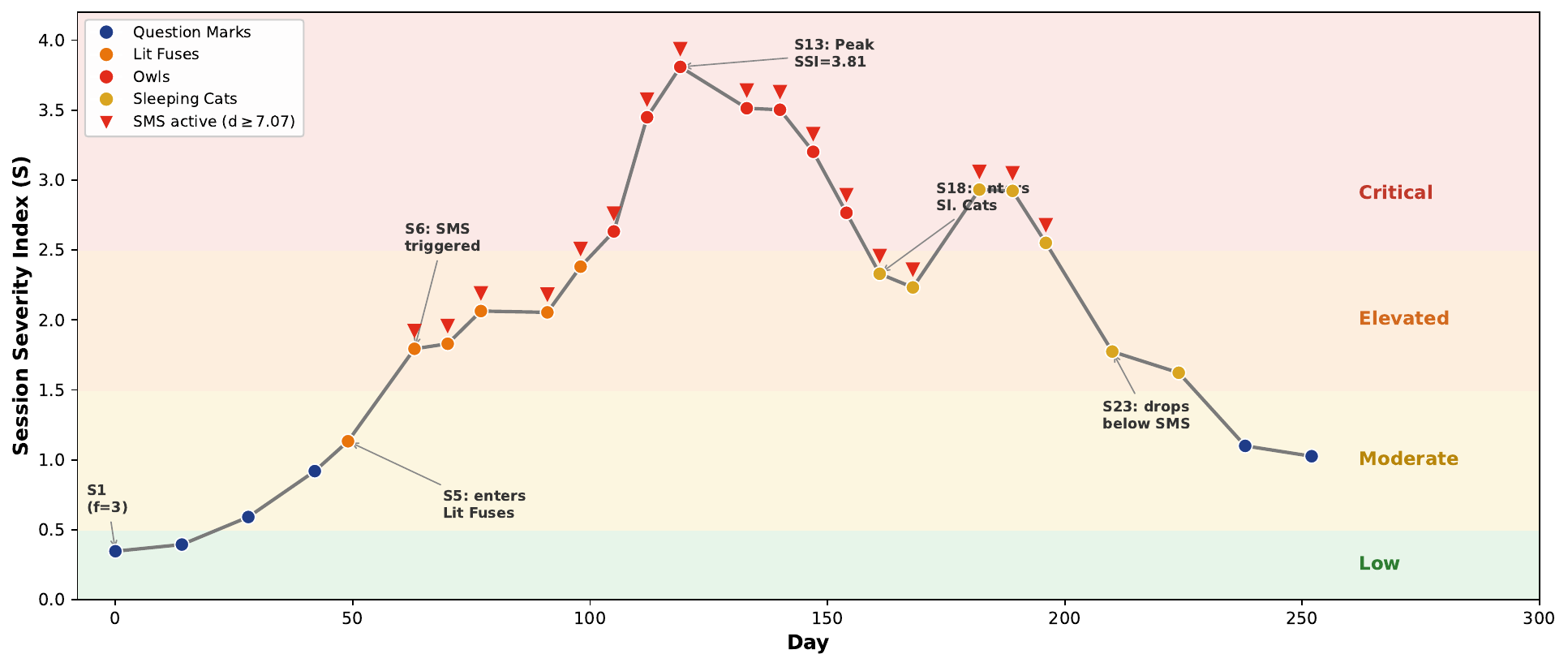}
\caption{Session Severity Index ($S$) over time for the Gas Fumes signal. Colored bands correspond to the SSI interpretation thresholds in Table~6. Node colors indicate the signal's current field region. The SSI peaks at Session~13 ($S = 3.81$) during the Owls phase and, while declining through the Sleeping Cats phase, briefly re-enters the Critical band at Sessions~20--22 ($S \approx 2.93, 2.92, 2.55$) even as $d$ drops below the SMS threshold.}
\label{fig:ssi}
\end{figure}

Fourth, the Session Severity Index ($S$) provides a complementary lens to the distance metric $d$ (Figure~\ref{fig:ssi}). The distance $d$ captures where the signal is at a given moment---a geometric snapshot of its current position on the field. The SSI weights that position by the cumulative frequency of occurrence $f$, reflecting how often the underlying event has actually been observed. An organization that tracks only $d$ might close signals prematurely once their coordinates improve; an organization that tracks only $S$ might overreact to high-frequency but low-severity nuisance signals. Together the two metrics give the full picture. For example, at Session~6 the SMS threshold triggers ($d = 9.89$) but the SSI is only 1.79---the position is alarming but the event has only been observed 12 times, suggesting the escalation may still be an emerging pattern rather than a confirmed crisis. By Session~24, $d$ has dropped to 5.63 (below the SMS threshold), but the SSI remains at 1.62 because 58 cumulative occurrences represent a substantial organizational history that should not be discounted simply because the current coordinates have improved.

Note: All 26 sessions follow the recommended biweekly cadence, yielding $\tau \approx 1.0$ at each update (except Sessions~5, 7, 8, 10--13, 15--19, 21, and 22 at $\tau = 0.5$ following one-week response intervals), demonstrating the model's behavior across both standard and rapid-response intervals.

\section{The WSCM as a Bridge to AI-Supported Analytics}

We designed the WSCM to be useful immediately, with no technology beyond a printed form and a pen. It is also designed to be the human layer of a more sophisticated analytical architecture that becomes available as organizations develop digital maturity.

The structured output of each cultivation session, a set of $(x, y)$ coordinate pairs with timestamps, associated observation metadata, and team notes, constitutes a time-series dataset in a standardized format. This dataset is directly compatible with machine learning pattern recognition. AI models trained on risk locus data across multiple signals and multiple organizations could, in principle, identify characteristic escalation trajectories, predict which signals in Question Marks are likely to reach Lit Fuses within a given timeframe, and surface organizational blind spots by comparing locus patterns against incident records.

The choice of a coordinate field over a categorical classification system follows directly from the intended use. Coordinates are continuous, mathematically tractable, and scale naturally to machine learning applications. Categories are not. A team working on a whiteboard and an AI system processing thousands of signals across organizations need to share the same data structure---coordinates provide that; categories do not.

\section{Discussion}

\subsection{What the Model Contributes}

We believe the WSCM contributes three things that do not currently exist in combination in any published framework. First, a coordinate-based representation of weak signals that preserves their continuous, dynamic character rather than forcing categorical classification. Second, the risk locus as an analytical concept, the idea that a risk's trajectory through a continuous field over time is the most information-rich representation of its organizational significance. Third, an explicit architectural bridge between human-generated field observation and AI-supported analytics, built into the model's design from the outset rather than retrofitted.

Individually, some of these ideas exist in adjacent literature. Risk matrices use two-dimensional representations. Time-series analysis of safety data is established. Human-AI collaboration in risk management is an active research area. What we introduce is a specific, practical, human-operated instrument that integrates all three. It is designed to work without technology but structured to feed technology when it becomes available.

\subsection{Current Scope and Next Steps}

This white paper presents the WSCM as a conceptual framework with an initial mathematical formalization. Several areas remain open for development. The NRS observation inputs rely on team judgment, which introduces subjectivity and potential for bias, including social desirability bias in near-miss reporting (Pfeiffer et al., 2010). The coordinate placement process we describe is deliberative rather than purely formula-driven, and refining its reproducibility across different teams is a priority. The risk locus is currently a qualitative concept; formalizing its statistical characterization is part of the planned methods follow-up.

This document does not include an empirical user evaluation of the WSCM. We chose to scope this document around the framework's architecture and mathematics, with empirical validation as the immediate next step. The architecture is sound, the mathematics are tractable, and the design is grounded in established theory from participatory design, behavioral nudging, and Safety-II. Evidence of real-world effectiveness will follow.

The pilot applications that informed this work involved limited sample sizes in healthcare and industrial construction. Generalizability across industries, organizational cultures, and risk types remains to be demonstrated. We expect the model to transfer to domains such as financial risk, cyber risk, or supply chain management, but this has not yet been tested.

\subsection{Design Implications}

Several design implications emerge from the framework and its pilot applications that we believe are relevant to practitioners and system designers considering adoption.

\textit{Start with paper, not software.} The WSCM was built to work without digital infrastructure. Organizations considering adoption should resist the urge to build an app first. A printed session form, a whiteboard, and a facilitator willing to run a thirty-minute meeting biweekly are sufficient. Digital tools should follow after the team has internalized the vocabulary and the session rhythm---typically after eight to twelve sessions.

\textit{The session is the intervention, not the field.} The cultivation field is a visualization tool, but the organizational value lives in the structured conversation that produces each position update. Designers building digital implementations should prioritize the session workflow---NRS elicitation, discussion capture, action assignment---over the visual display of the field.

\textit{Quiet sessions are not empty sessions.} When no signals move for several consecutive sessions, teams may perceive the tool as generating no value. System designers should build in confirmatory feedback: a dashboard element that explicitly acknowledges stable signals as evidence that controls are holding, not as absence of information.

\textit{Design for the committee, not the individual.} The model's $n$-cap mechanism reflects a core design principle: collective assessment is more reliable than individual assessment. Digital implementations should make it easy for multiple team members to contribute independent NRS scores before seeing each other's ratings, preventing anchoring bias.

\textit{Preserve the locus.} The risk locus---the trajectory a signal traces across the field, is the most analytically valuable output of the WSCM. Any implementation, paper or digital, must preserve the full positional history of every signal. Implementations that show only the current position discard most of the model's organizational intelligence.

\subsection{Roadmap}

Three lines of future work follow directly from this document. First, a methods paper formalizing the coordinate computation (see Section~3.5 for the initial formalization), translating the NRS observation inputs into $(x, y)$ coordinates through an explicit, validated formula system, including handling of irregular observation intervals and the reporting culture correction. This paper is in preparation.

Second, a prospective pilot study deploying the WSCM in at least two organizational settings across a minimum 12-month observation period, tracking signal evolution, intervention decisions, and safety outcomes. This study will provide the empirical validation that the current paper does not claim.

Third, a technical specification for AI integration, defining the data pipeline from WSCM session outputs to machine learning analysis, including the pattern recognition models, feedback mechanisms, and human-override protocols required for responsible deployment. This work will proceed in parallel with the pilot study.

\section{Conclusion}

Organizational safety fails at the last mile, the space between what frontline workers know and what organizational systems act on. The WSCM closes this gap. It gives weak signals a structured home, a shared vocabulary, and a way to track how they change over time. It asks nothing of organizations beyond regular attention and honest observation. What it produces is rare: a living record of risks that are still forming, not just the ones that have already arrived.

The model does not promise certainty. Weak signals are by their nature uncertain and periodically manifest as anomalies in otherwise routine work. What the WSCM offers is a discipline, a practice of collective attentiveness that makes early detection a repeatable organizational capability rather than an occasional stroke of luck. In safety-critical domains, the difference between these two is not academic. It is the difference between catching the lit fuse before it reaches the powder and explaining afterward why the warning signs were missed.

We introduce this framework as a first, deliberate step. The conceptual architecture is established here. The mathematical formalization, empirical validation, and AI integration follow. Each step depends on the one before it, and this paper is the one before all of them.

Organizations interested in piloting the WSCM can begin today. Section~4.4 describes the minimum viable deployment: a team of four to eight people, a printed session form, and thirty minutes of facilitated discussion every two weeks. No software required. Start there, and see what your team begins to notice.


\section*{Author Contributions and AI Use Disclosure}

\textbf{Author contributions.} Maurice Codourey conceived the Weak Signal Cultivation Model, developed the theoretical framework, designed the coordinate field architecture and quadrant taxonomy, conducted the pilot applications, and is responsible for all conceptual and empirical content of this paper. Emmanuel A.\ Gonzalez contributed to the mathematical formalization of the coordinate derivation model, the movement scoring system, and the scientific refinement of the computational framework.

\textbf{AI use disclosure.} All ideas, theoretical contributions, model design, and research findings in this paper originate from the human authors. Claude (Opus 4 and Opus 4.6, Anthropic) was used to draft and refine academic prose from the authors' outlines and instructions, generate Python code for figure production, recompute the illustrative simulation, and produce LaTeX typesetting for tables and document formatting. Claude (Sonnet 4, Anthropic) was used in earlier iterations to compute initial numerical outputs for the illustrative simulation in Section~5 from author-specified inputs and scenarios; the simulation data are not derived from field observation. For this revised version (v2), Claude (Fable~5, Anthropic) was used to diagnose internal-consistency, citation, and numerical errors across the manuscript, and Claude (Opus~4.8, Anthropic) was used to verify those findings against source, recompute the worked examples and the Gas Fumes trajectory, and regenerate the figures. Grammarly was used for grammar and style review. Perplexity was used to support literature search and reference verification. The authors reviewed, revised, and approved all output as the final step before submission. No AI tool contributed to the conceptual design, theoretical framework, or analytical conclusions of this work. The authors take full responsibility for the accuracy, integrity, and originality of all content.


\section*{References}
\vspace{4pt}

\begin{sloppypar}
\begin{list}{}{\setlength{\leftmargin}{1.5em}\setlength{\itemindent}{-1.5em}\setlength{\itemsep}{4pt}\setlength{\parsep}{0pt}}
\item Ansoff, H. I. (1975). Managing strategic surprise by response to weak signals. \textit{California Management Review}, 18(2), 21--33.
\item Aboagye-Nimo, E., Raiden, A., King, A., \& Tietze, S. (2015). Using tacit knowledge in training and accident prevention. \textit{Proceedings of the Institution of Civil Engineers --- Management, Procurement and Law}, 168(5), 232--240.
\item Bj{\o}rn, P., \& {\O}sterlund, C. (2014). \textit{Sociomaterial-Design: Bounding Technologies in Practice}. Springer.
\item Codourey, M. (2025). Weak signal farming quadrant: A human-centric approach for frontline risk detection and resilience. ResearchGate. \url{https://www.researchgate.net/publication/395334608}
\item Dekker, S. (2014). \textit{The field guide to understanding `human error'} (3rd ed.). CRC Press.
\item Eraut, M. (2004). Informal learning in the workplace. \textit{Studies in Continuing Education}, 26(2), 247--273.
\item Farinha, J., Vesnic-Alujevic, L., \& P\'{o}lvora, A. (2023). Scanning deep tech horizons---Participatory collection and assessment of signals and trends. Publications Office of the European Union. \url{https://doi.org/10.2760/48442}
\item Gnoni, M. G., \& Saleh, J. H. (2017). Near-miss management systems and observability-in-depth: Handling safety incidents and accident precursors in light of safety principles. \textit{Safety Science}, 91, 154--167.
\item Gonzalez, E. A., Presto, R. S., Remacha, A. C., \& Santos, A. N. (2015). A model describing hazard identification effectiveness of workers in the construction and maintenance industry. In \textit{Proceedings of the 8th IEEE GCC Conference and Exhibition}, Muscat, Oman. \url{https://doi.org/10.1109/IEEEGCC.2015.7060044}
\item Hasdell, P. (2016). Participatory design: Re-evaluation as a socio-material assembly. \textit{Ethnographic Praxis in Industry Conference Proceedings}, 313--326.
\item Henderson, B. D. (1970). \textit{The product portfolio}. Boston Consulting Group.
\item Hollnagel, E. (2014). \textit{Safety-I and Safety-II: The past and future of safety management}. Ashgate Publishing.
\item Hollnagel, E., Woods, D. D., \& Leveson, N. (2006). \textit{Resilience Engineering: Concepts and Precepts}. Ashgate Publishing.
\item Kletz, T., \& Amyotte, P. (2019). \textit{What went wrong? Case histories of process plant disasters and how they could have been avoided} (6th ed.). Elsevier.
\item Reason, J. (1997). \textit{Managing the risks of organizational accidents}. Ashgate Publishing.
\item Snowden, D. J., \& Boone, M. E. (2007). A leader's framework for decision making. \textit{Harvard Business Review}, 85(11), 68--76.
\item Thaler, R. H., \& Sunstein, C. R. (2021). \textit{Nudge: The final edition}. Penguin Books.
\item Tversky, A., \& Kahneman, D. (1974). Judgment under uncertainty: Heuristics and biases. \textit{Science}, 185(4157), 1124--1131.
\item Poon, A., Luebke, M., Loughman, J., Lee, A., Guerrero, L., Sterling, M., \& Dell, N. (2023). Computer-mediated sharing circles for intersectional peer support with home care workers. In \textit{Proceedings of the ACM on Human-Computer Interaction}, 7(CSCW1), Article 39. \url{https://doi.org/10.1145/3579472}
\item Pfeiffer, Y., Manser, T., \& Wehner, T. (2010). Conceptualising barriers to incident reporting: A psychological framework. \textit{Quality and Safety in Health Care}, 19(6), e60.
\item United Nations Children's Fund (UNICEF). (2019). \textit{Human-centred design in the field}. UNICEF Office of Innovation. \url{https://www.unicef.org/innovation/reports/human-centred-design-field}
\item Van der Schaaf, T. W. (1992). \textit{Near miss reporting in the chemical process industry}. Eindhoven University of Technology.
\item Varanasi, U.\,S., Leinonen, T., Sawhney, N., Tikka, M., \& Ahsanullah, R. (2023). Collaborative sensemaking in crisis: Designing practices and platforms for resilience. In \textit{Proceedings of the 2023 ACM Designing Interactive Systems Conference (DIS '23)}, 2537--2550. \url{https://doi.org/10.1145/3563657.3596093}
\item Vaughan, D. (1996). \textit{The Challenger launch decision: Risky technology, culture, and deviance at NASA}. University of Chicago Press.
\item Weick, K. E. (1995). \textit{Sensemaking in organizations}. Sage Publications.
\item Wicht, P. (2025). Weak signal analysis. Whispers \& Giants. \url{https://www.whispersandgiants.com/2025/02/20/weak-signal-analysis/}
\end{list}
\end{sloppypar}


\appendix
\appendixsectionformat

\section{Technical Reference: Parameter Defaults and Derivations}

This appendix documents how each default value in Table~\ref{tab:params} was determined. Parameters fall into three categories: mathematically derived values that follow directly from a specified design target; simulation-calibrated values tuned to produce consistent behavior across the two worked signals (Gas Fumes, Section~5; and Complacency, presented in Codourey, 2025); and design judgments grounded in operational or organizational reasoning. All three categories are reported transparently so that practitioners adapting the model to different contexts understand which values are fixed by logic and which are open to tuning.

\subsection{$T_{\text{ref}} = 14$ days --- Deployment Convention}
$T_{\text{ref}}$ is the organizational reference period used to compute the dimensionless time parameter $\tau = \Delta t / T_{\text{ref}}$. Its default value of 14 days matches the recommended biweekly cultivation session cadence directly: at normal cadence, every session produces $\tau = 1.0$, and all update formulas operate at their intended design point. $T_{\text{ref}}$ is the single parameter an organization changes when adopting a different session frequency---weekly deployment sets $T_{\text{ref}} = 7$; monthly sets $T_{\text{ref}} = 30$. All other parameters remain unchanged.

\subsection{$\mu = 0.087$ --- Derived from Half-Life Specification}
The $y$-decay parameter $\mu$ governs how quickly Risk Growth Potential diminishes in the absence of new reports (Equation~11). Its value is derived directly from a target half-life of eight reference periods. Setting the half-life condition:
\[
0.5 = e^{-\mu \cdot 8} \implies \mu = \frac{\ln 2}{8} = \frac{0.6931}{8} \approx 0.0866 \approx 0.087
\]
At the default biweekly cadence, eight reference periods equals 16 weeks (approximately four months). This target was chosen to reflect the organizational judgment that a signal reported only once should retain meaningful growth potential for roughly one quarter before fading---long enough to reappear if conditions recur, short enough that genuine dormancy is eventually reflected in the coordinates.

\subsection{$\lambda = \nu = 0.75$ --- Calibrated to 53\% Weight at Normal Cadence}
The time-sensitivity parameters $\lambda$ (for $x$) and $\nu$ (for $y$) control how rapidly the base recency weights saturate toward their ceilings (Equations~3--4). Their default of 0.75 produces a specific update weight at the normal session cadence ($\tau = 1.0$):
\[
\alpha(1.0) = \alpha_{\text{base}} \cdot (1 - e^{-0.75}) = 0.90 \times 0.5276 \approx 0.475
\]
This means a new assessment at normal cadence moves the coordinate approximately 47.5\% of the distance toward the new value, before the momentum term $\kappa$ is added. The design intent is that no single session should dominate the signal's history, while rapid repositioning remains possible when consecutive sessions accumulate. At a longer gap ($\tau = 2$): $\alpha(2.0) \approx 0.70$; at a rapid follow-up ($\tau = 0.5$): $\alpha(0.5) \approx 0.28$.

\subsection{$\alpha_{\text{base}} = \beta_{\text{base}} = 0.90$ --- Simulation-Calibrated Ceiling}
The recency weight ceilings were calibrated across the Gas Fumes (26-session, Section~5) and Complacency (Codourey, 2025) simulations to satisfy three conditions simultaneously: (1)~a signal at peak concern can escalate from entry within 10--13 sessions; (2)~de-escalation after effective intervention is visible within 4--6 sessions; and (3)~the $n$-cap mechanism at small committee size ($n = 1$) does not prevent meaningful repositioning. The value 0.90 is the minimum ceiling satisfying all three conditions in both simulations. Values below 0.80 produced sluggish escalation; values above 0.95 allowed single sessions to destabilize established loci.

\subsection{$\delta = 0.50$, $\eta = 0.30$, $\phi = 0.30$ --- Momentum Weight Allocation}
The three terms in the consensus momentum formula $\kappa$ (Equation~5) are weighted by $\delta$ (magnitude of change), $\eta$ (sustained directional agreement), and $\phi$ (reporter volume). Their defaults reflect a deliberate ordering of evidence strength: $\delta$ is highest because a large jump in the new assessment is the strongest single indicator that conditions have genuinely changed; $\eta$ and $\phi$ are equal secondaries capturing persistence and committee size respectively.

The values were calibrated to ensure $\kappa \leq 1$ under any realistic input combination. At maximum inputs ($|\Delta x| = 10$, $k = k_{\text{ref}}$, $n = n_{\text{ref}}$):
\[
\kappa_{\max} = \min\!\left(1,\; 0.50 \cdot \frac{10}{10} \cdot \frac{k_{\text{ref}} + 1}{k_{\text{ref}}} + 0.30 \cdot 1 + 0.30 \cdot 1\right) = \min(1,\; 0.60 + 0.30 + 0.30) = \min(1,\; 1.20) = 1.0
\]
Combined with $\alpha_{\text{eff}} = \min(n_{\text{cap}}, \alpha(\tau) + \kappa) \leq 1$, the position update always remains a valid convex combination.

\subsection{$\psi = 0.50$ --- Reversal Amplifier}
When a team reverses its directional assessment and the committee is large ($n \geq n_{\text{ref}}$), the effective weight is multiplied by $\rho = 1 + \psi \cdot n_{\text{boost}}$. At $\psi = 0.50$ and full committee: $\rho = 1.5$, meaning a confident large-committee reversal repositions the signal 50\% faster. The value was chosen so that a full reversal is visible within two to three sessions rather than five to six. Values below 0.30 produced imperceptibly slow reversals; values above 0.70 created instability in signals oscillating near a quadrant boundary.

\subsection{$k_{\text{ref}} = 5$ --- Sustained-Agreement Horizon}
$k_{\text{ref}}$ defines the number of consecutive same-direction sessions required for full momentum effect. At biweekly cadence, five sessions spans ten weeks, approximately one organizational quarter. A team consistently reporting the same directional shift over a full quarter provides strong evidence of a genuine trend rather than noise. Below $k_{\text{ref}}$, momentum accumulates proportionally ($k/k_{\text{ref}}$), so momentum builds gradually rather than switching on abruptly.

\subsection{$n_{\text{ref}} = 5$ --- Minimum Viable Committee}
Five reporters reflects the minimum viable team size for meaningful group consensus in occupational safety practice---consistent with small-group deliberation norms in Safety-II and human-factors research. Below five, $n_{\text{boost}} = n/n_{\text{ref}} < 1$ and both the $\phi$ term in $\kappa$ and the $n$-cap ceiling are reduced proportionally, preventing under-resourced sessions from driving large positional shifts.

\subsection{$\alpha_{\min} = 0.70$ --- Solo-Reporter Floor}
The $n$-cap floor sets the minimum effective ceiling when only one reporter is present:
\[
n_{\text{cap}}(n{=}1) = 0.70 + (1 - 0.70) \cdot \frac{1}{5} = 0.70 + 0.06 = 0.76
\]
A solo reporter can therefore move a signal by up to 76\% of the distance to the new assessment per session---responsive enough to capture a genuine first observation, constrained enough to prevent single-reporter noise from destabilizing an established risk locus. This is verified in the Gas Fumes simulation: at Session~2 ($n = 1$), $\alpha_{\text{eff}} = 0.535 < 0.76$, confirming the cap operates as intended.

\subsection{$y_{\min} = 0.50$ --- Decay Floor}
The passive $y$-decay floor prevents a signal from disappearing silently while it remains open in the register. On the $[0, 10]$ field, $y_{\min} = 0.50$ represents 5\% of the axis range---low enough that a genuinely de-escalating signal approaches it without distortion, high enough that the signal retains a minimal growth-potential reading prompting a team decision to actively retire it rather than letting it drift unnoticed.

\subsection{$d_{\text{close}} = 1.0$ --- Closure Threshold}
A signal may only be retired when its Euclidean distance from the origin falls below $d_{\text{close}} = 1.0$. On the $[0, 10]$ field this corresponds to coordinates near $(0.70, 0.70)$, since $\sqrt{0.70^2 + 0.70^2} \approx 0.99 < 1.0$. The threshold ensures closure requires genuinely low concern on both dimensions simultaneously, preventing closure when only one axis has dropped. The value 1.0 is the smallest round number enforcing this dual-axis condition without being so restrictive that well-managed signals cannot be retired.

\subsection{SMS Escalation Threshold $d \geq 7.07$ --- Geometric Derivation}
The Safety Management System escalation threshold is derived directly from the field geometry:
\[
d_{\text{SMS}} = \sqrt{5^2 + 5^2} = \sqrt{50} \approx 7.07
\]
This is the Euclidean distance from the field origin $(0, 0)$ to the field centre $(5, 5)$---the intersection of the quadrant boundaries. Escalation therefore triggers when a signal's combined risk position surpasses the geometric centre of the field. Because the trigger is based on Euclidean distance from the origin, it can also be reached by a single dominant dimension: a signal at $(9.51, 2.71)$, for instance, escalates from within Lit Fuses before its growth-potential coordinate reaches the midpoint. The threshold is fully determined by the field dimensions and requires no additional calibration.

\vspace{14pt}
\noindent\rule{\columnwidth}{0.4pt}
\vspace{4pt}
\noindent{\footnotesize\textcopyright\ Codourey \& Gonzalez, 2026. All rights reserved.\hfill Version 2.0 --- June 2026}

\end{document}